# A transition from antiferromagnetism to spin-glass freezing in single crystalline, $Dy_2PdSi_3$

K.K. Iyer[+], P.L. Paulose[+], E. V. Sampathkumaran[+], H. Bitterlich[*], G. Behr[*] and Löser[*]

[+]Tata Institute of Fundamental Research, Homi Bhabha Road, Mumbai – 400 005, India
[*]Leibniz-Institut für Festkörper- und Werkstoffforschung Dresden, Postfach 270116, D-01171 Dresden, Germany

*Abstract*

*We report low temperature (1.8-15 K) ac magnetic susceptibility and isothermal remanent magnetization behavior for the single crystalline form of $Dy_2PdSi_3$, crystallizing in a $AlB_2$-derived hexagonal structure, for two orientations. The results, while confirming the existence of two magnetic transitions, one at 8 K and the other at 2.5 K, as inferred from previous dc magnetization studies, suggest that the 2.5K-transition is spin-glass-like, with an interesting sensitivity to the application of an external dc magnetic field. Thus, this compound appears to undergo a transition from antiferromagnetism to spin-glass freezing as the temperature is lowered.*

## INTRODUCTION

The ternary intermetallic compounds of the type, $R_2PdSi_3$ (R= Rare-earth), crystallizing in a $AlB_2$-derived hexagonal structure [1] exhibit anomalous magnetic properties [See, for instance, Refs. 2-5, and articles cited therein]. Many of the heavy R members have been shown to exhibit two magnetic transitions, and the efforts to understand the origin of these have just begun in the literature [4,5].

In this article, we focus on the compound $Dy_2PdSi_3$ which turned out to be one of the few R-based intermetallic compounds reported to exhibit giant magnetoresistance at low temperatures several years ago [2]. The onset of antiferromagnetic ordering takes place at 8 K in this compound, while initial dc magnetization studies indicated the existence of a second transition around 2.5 K [2]. The 2.5-transition behaves like a spin-glass in ac magnetic susceptibility ($\chi$) of polycrystals [4]; this is a rare phenomenon among stoichiometric materials, presumably arising from a small degree of crystallographic disorder thereby enabling randomization of the exchange coupling. If so, this is one of the few R-based compounds exhibiting re-entrant magnetism (in the following sense: paramagnetic to antiferromagnetic to spin-glass-like). It is therefore important to characterize the magnetic behavior of this compound at low temperatures in the single crystalline form, particularly focusing on the spin-glass features. We have earlier reported dc magnetization behavior on the single crystals of this compound [3] and we report here the results of ac $\chi$ and isothermal remanent magnetization ($M_{IRM}$) measurements on single crystals at low temperatures keeping this objective in mind.

## EXPERIMENTAL DETAILS

The single crystal pieces employed in the present investigations are the same as those in Ref. 3. Ac $\chi$ measurements below 15 K were performed for two orientations $H_{ac}//[0001]$ and $H_{ac}//[10\bar{1}0]$ at various frequencies ($\nu$) with an ac magnetic field (H) amplitude of ($H_{ac}=$) 1 Oe, employing a commercial superconducting quantum interference device (Quantum Design), in the absence as well as in the presence of dc magnetic fields (1 and 15 kOe) parallel to the direction of $H_{ac}$. The time-dependent $M_{IRM}$ measurements were performed after cooling the sample in zero dc magnetic field to a required temperature, followed by switching on the dc H of 5 kOe for 5 mins; the decay of M was then tracked as a function of time ($t$) after switching off the dc H.

## RESULTS AND DISCUSSION

The results of ac $\chi$ measurements are shown in figures 1 and 2. Apart from the fact that there is a weak anisotropy in the absolute values of ac $\chi$ in the magnetically ordered state, the following observations are made in these data.

For both the orientations, in zero dc field, (i) there is a peak or flattening in the real part ($\chi'$) of ac $\chi$ around 8 K and the position of these features are essentially independent of $\nu$. There is no prominent feature in the imaginary part ($\chi''$) of ac $\chi$ at 8 K (see the curves for $\nu= 1.2$ Hz). These findings are consistent with the proposal of a long-range magnetic order with a well-defined magnetic structure (say, antiferromagnetic [2-4]). (ii) There is a peak in $\chi'$ at 2.5 K for $\nu= 1.2$ MHz, which shifts to a higher temperature with increasing $\nu$, for instance, by about 1 K for $\nu= 1222$ Hz, as though this peak is merging with the profile of the 8K-transition; correspondingly, there is a prominent feature in $\chi''$ also and this $\nu$-dependence is more obvious from the upward-temperature shift of $\chi''$-curves with increasing $\nu$. These findings suggest that the 2.5K-transition could be of a spin-glass-type. However, the observed magnitude of the shift of the peak temperature is large, untypical of canonical spin-glasses, as noted for polycrystals as well [4].

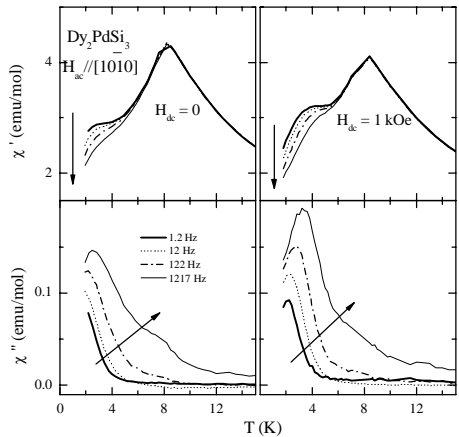

**Fig. 1:** Ac susceptibility behavior of $Dy_2PdSi_3$ at various frequencies for the orientation $H_{dc}//H_{ac}//[10\bar{1}0]$. The curves move with increasing ν in the direction of the arrows.

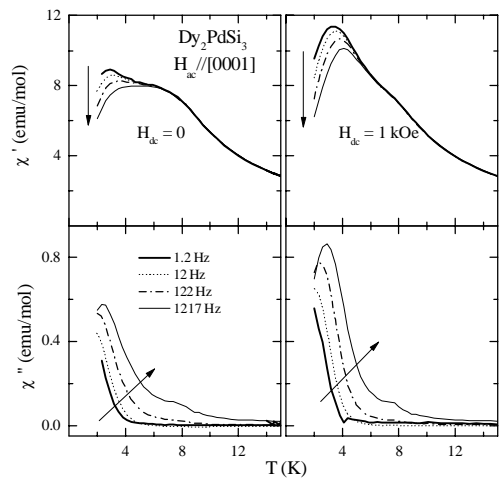

**Fig. 2:** Ac susceptibility behavior of $Dy_2PdSi_3$ at various frequencies for the orientation $H_{dc}//H_{ac}//[0001]$. The curves move with increasing ν in the direction of the arrows.

The relative intensity of the 2.5K-peak of χ'-curve for a given ν with respect to that of the 8K-peak is smaller for $H_{ac}//[10\bar{1}0]$ while compared with the respective ratio for $H_{ac}//[0001]$. We now point out an interesting observation on the influence of the application of dc H on this feature. From the absolute values of χ', it appears that, for H= 1 kOe, the intensity of 2.5K-feature goes up for both the orientations. However, this increase is very dramatic for H//[0001]. This is an unusual finding, the origin of which is not clear. Further increase of dc H to 15 kOe washes out all the features in the ac χ data (and hence not shown in the form of a figure), thereby implying a destruction of spin-glass freezing in favor of a well-defined long-range magnetic structure, which is consistent with isothermal magnetization behavior discussed in Ref. 3.

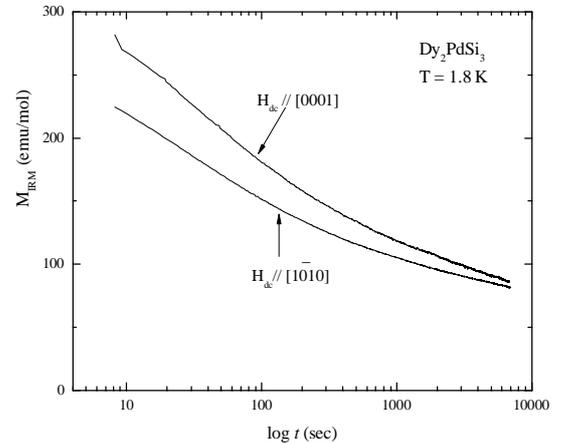

**Fig. 3:** Time dependent isothermal remanent magnetization behavior at 1.8 K just after the field (applied with respect to the orientation specified in the figure) was switched off.

$M_{IRM}$ data obtained at 1.8 K are shown in Fig. 3 and it is clear that there is a slow decay $M_{IRM}$, typical of spin-glasses. The actual functional form of $M_{IRM}(t)$ appears to be complicated, though it is logarithmic to start with. Such complicated forms of $M_{IRM}(t)$ have often been reported in the spin-glass field.

## CONCLUSION

The present results on single crystals of $Dy_2PdSi_3$ suggest that this compound is characterized by paramagnetic to antiferromagnetic to spin-glass transitions as the temperature is lowered – a finding not so common among stoichiometric rare-earth compounds.